\documentclass[10pt,twocolumn,letterpaper]{article}

\usepackage{iccv}
\usepackage{times}
\usepackage{epsfig}
\usepackage{graphicx}
\usepackage{amsmath}
\usepackage{amssymb}
\usepackage[misc,geometry]{ifsym}

\usepackage[breaklinks=true,bookmarks=false]{hyperref}

\iccvfinalcopy 

\def\iccvPaperID{****} 

\ificcvfinal\fi

\begin{document}

\title{COVID Detection in Chest CTs: Improving the Baseline on COV19-CT-DB}

\author{
Radu Miron\\

\and
Cosmin Moisii\\

\and
Sergiu Dinu\\

\and
Mihaela Breaban  (\Letter)

\and
SenticLab, Iasi, Romania\\
Faculty of Computer Science, "Alexandru Ioan Cuza" University of Iasi, Romania\\
pmihaela@info.uaic.ro


}

\maketitle
\ificcvfinal\thispagestyle{empty}\fi

\begin{abstract}
The paper presents a comparative analysis of three distinct approaches based on deep learning for COVID-19 detection in chest CTs. The first approach is a volumetric one, involving 3D convolutions, while the other two approaches perform at first slice-wise classification and then aggregate the results at the volume level. The experiments are carried on the COV19-CT-DB dataset, with the aim of addressing the challenge raised by the MIA-COV19D Competition within ICCV 2021. Our best results on the validation subset reach a macro-F1 score of 0.92, which improves considerably the baseline score of 0.70 set by the organizers.
\end{abstract}

\section{Introduction}
There is a high effervescence in the AI community trying to provide enhanced tools to assist the medical diagnosis, with a special focus on medical imaging, exploiting but also triggering important advancements in the area of computer vision and deep learning. The pandemic scenario we face today clearly calls for such approaches in order to be able to address the high incidence rate which overwhelms the medical system. Medical imaging, mostly in the form of x-Rays and CTs, is used to assess the lung involvement. In this context, COVID-19 datasets are publicly released and competitions are organized, aiming at involving and stimulating the AI community to produce models that can accurately detect Covid-19 affections in the lungs.

The current work addresses the challenge raised by the MIA-COV19D competition\footnote{https://mlearn.lincoln.ac.uk/mia-cov19d/} organized within the ICCV 2021 conference. The challenge consists in tackling a two-class classification problem on the COV19-CT-DB dataset consisting of CTs classified in two groups: COVID patients and non-COVID patients, with the second class containing both healthy patients or patients presenting lung lesions due to other causes. The dataset was split by the organizers into 3 subsets: training, validation and test, with the first two subsets also exposing the class label. The baseline set for this classification task by the organizers, for comparison purposes, is a 0.70 macro-F1 score on the validation set, obtained with a CNN-RNN network described in \cite{kollias2021mia}, which is the result of authors' previous work reported in \cite{kollias2018deep,kollias2020deep, kollias2020transparent}. The final evaluation of the models built by the participants in the competition will be made on the test set that was released with no class/label information.

Given the nature of the data, we address the challenge in two different ways: 1) treating the CT as a volume and thus using 3D convolutions, and 2) treating the CT as a set of 2d images (slices), the second approach calling for 2D convolutions and predictions at slice level, followed by an aggregation step where the outputs at slice level are aggregated into a response at CT level. 

Our second approach is motivated by the results we obtained in the 2020 and 2021 ImageClefMed competitions on tuberculosis tasks, where our team was ranked the first \cite{senticlabImageClef2020, senticlabImageClef2021}. This approach needs in the training phase more refined information: it needs CT slices split in two categories: slices that present COVID lesions and slices that do not present COVID lesions (meaning healthy or other affection types). Because the dataset exposes this information only at CT level and a COVID case presents both healthy and affected lung slices, we needed to come up with a procedure to extract COVID-affected slices from the COVID CTs in the training set. We use at first a naive method to label slices, where all the slices belonging to a COVID CT are considered as slices presenting COVID lesions. As a second more refined method, we use a model we had trained for the tuberculosis classification task in the ImageClefMed 2021 competition that is able to detect slices presenting different types of tuberculosis lesions. Since the ImageClef data does not contain any COVID patients, we expect that the results are not highly accurate in identifying slices presenting COVID affections and this approach also introduces noise, mainly as slices presenting other affections and not COVID, because these could be present in both COVID and non-COVID CTs. After slice labeling, we use two distinct approaches for COVID classification: one that makes use of a manually designed NN architecture working with "mini-volumes" made of 3 slices and one that performs NN architecture search with sharpDARTS\cite{hundt2019sharpdarts}.

The paper is structured as follows. Section \ref{sec:data} describes the COV19-CT-DB dataset. Section\ref{sec:volumetric} describes our first approach that makes inference directly at the CT level. Sections \ref{grouped_slices} and \ref{sec:NNSearch} describe the two approaches that learn to classify the slices as COVID/non-COVID and then aggregate the results at CT level. Section \ref{sec:experiments} presents the results obtained on the training and validation sets and section \ref{sec:conclusions} concludes the paper.

\section{The dataset}\label{sec:data}

The only dataset used in the experimental analysis is COV19-CT-DB \cite{kollias2021mia}, provided in the MIA-COV19D competition. 

Our slice-based approaches described in sections \ref{grouped_slices} and \ref{sec:NNSearch} do not involve directly an external dataset, but only a pretrained model built by us in the ImageClef2021 competition on CT data consisting of non-COVID patients that present tuberculosis lesions; this pretrained model serves only to select a subset of slices from the COVID class in the COV19-CT-DB training dataset. We emphasise the fact that using this pretrained model built actually for a tuberculosis task and used here to select slices presenting COVID might introduce noise in the training set we build at slice level.

The provided COV19-CT-DB training set has a total of 1560 CT scans. The class distribution is 690 COVID-19 cases versus 870 Non-COVID19 cases. Out of these, we found 8 volumes with less than 20 slices and 5 volumes with more than 700 slices in their corresponding directories. 

The provided validation set has a total of 374 CT scans. The class distribution is 165 COVID-19 cases versus 209 Non-COVID-19 cases. Out of these, we found no volume with less than 20 slices and 1 volume with more than 700 slices in their corresponding directories.  

As explained in \cite{kollias2021mia}, the labeling of the dataset was made by a consensus of two radiologists and two pneumologists. The difference with other publicly available datasets is the annotations based on expert opinion rather than just positive RT-PCR testing. 

The whole COV19-CT-DB database consists of about 5000 CT scans, corresponding to more than 1000 patients and 2000 subjects, leading to 3455 cases for the test set.

The images are provided through jpg format by  clipping the voxel intensities using a window/level of 350 Hounsﬁeld units (HU) and 1150 HU and normalization to the range of [0,1] of each slice from a whole volume rather than the dicom files. 

Due to the large number of slices per volume and large dimensions of an individual slice, in our approaches we usually rescale the given images. Each classification approach described in the sections below will also list the modifications done to the input data.

\section{A volumetric approach}\label{sec:volumetric}

Convolutional neural networks have emerged as a successful tool in tackling a wide range of image vision tasks. These architectures have obtained, ever since the appearance of Alex-Net \cite{krizhevsky2012imagenet}, state of the art results on tasks like image classification, object detection and segmentation.   

Even though applications in 2D image tasks have shown great results, architectures for 3D image tasks still have to prove themselves as good contenders for state of the art results.

Nowadays benchmarks use pretraining for improving results. Since large 3D datasets are not so numerous, we turn our attention to a specific kind of model to fully benefit of the advantages of pretrained models.

We use an inflated convolutional neuronal network pretrained on Kinetics dataset \cite{inflated2017}. Inflated convolutions are obtained by expanding filters and pooling kernels of 2D ConvNets into 3D, resulting in the possibility to learn spatio-temporal feature extractors from 3D images while using successful ImageNet architectures. Due to gpu limitations we only used an Inflated ResNet50 model \cite{resnet2015}. 

For capturing long range dependencies within the slices of a same volume, we use non-local features which have been proven to increase the results of basic architectures. \cite{nonlocal2017}

The final model is an Inflated 3D ResNet50 with non local operations on the second and third layers (based on the official implementation from PyTorch). \footnote{https://pytorch.org/vision/0.8/\_modules/torchvision/models/resnet.html} 

Due to large number of parameters, 3D conv nets are easy to overfit. In order to mitigate the overfitting, we use label smoothing \cite{labelsmoothing} with cross entropy and Sharpness Aware Minimization \cite{sam} on top of SGD as loss function and optimizer, respectively. We also use augmentations like Random Horizontal and Vertical Flip, Gaussian Blur and Contrast and Color changes. We also tried with Cut-Out and Affine transformations, but the results were not improved compared to basic augmentations mentioned above. We may also flip the volume on the depth axis. All these augmentations are done with 0.5 chance.

The training procedure lasts for 150 epochs. Initial learning rate is set to 1e-3 and the learning rate scheduler is cosine scheduler with 5 epochs of warm-up. Batch size is 2.

The input consists of volumes of size $128 \times 224 \times 224$. First, we resize all the images to $224 \times 224$ and take only $128$ slices from each volume (with padding if necessary). Since there are large volumes in our dataset, we solve this issue using the following rule: if a volume has between $128k$ and $128(k+1)$ slices, we choose the starting slice a random number between $0$ and $k$ and sample every $k^{th}$ slice, discarding the others, generating a volume of 128 slices. This way, we make sure that most of the volume is preserved. Our procedure i based on the fact that volumes with large number of slices (corresponding to small slice thickness) admit a transversing with a small sliding window without great loss of information for covid diagnosis.

In order to support the results we obtained with the validation dataset provided by the organizers, we also created another 4 folds and completed for each fold the training and validation steps.

During the inference process, parts of a single volume will be several times input for the model. If a volume has between $256k$ and $256(k+1)$ slices, we choose as starting slice each number between $0$ and $k$ and take each $k^{th}$ slice discarding the others. This procedure generates $k + 1$ sub-volumes of 256 slices at inference time. For each sub-volume we apply horizontal or vertical flip or depth axis flip. This means $8(k+1)$ inferences for a certain volume. For inference we define two important thresholds: the confidence threshold for non-covid label and confidence threshold for both labels. Every non-covid prediction which is below the first threshold is ignored. The same happens for the second threshold but with both of the labels. After the eliminations based on the thresholding, the most frequent diagnosis is given to the specific volume. 

For the final prediction pipeline, we constructed an ensemble with 2 models from each fold that we trained on, stored at the epochs we obtained the highest validation accuracy on.  

Some of the results of folds $2 - 5$ are mentioned in the table below:

\begin{table}[h!]
\centering
 \begin{tabular}{||c c||} 
 \hline
 Fold/Epoch & F1 score \\ [0.5ex] 
 \hline\hline
  $2 / 93$ & $0.936$ \\ 
  $3 / 99$ & $0.923$ \\
  $4 / 99$ & $0.936$ \\
  $5 / 96$ & $0.925$ \\
  official / ensemble & $0.9234$ \\ [1ex] 
 \hline
 \end{tabular}
\end{table}

\section{A slice based approach with mini-volumes}\label{grouped_slices}
This method is a slice based approach with the addition that it incorporates the immediate previous and next slice creating a "minivol". It uses both slice and lung (side) level labels which we obtained in two steps:

\begin{itemize}
    \item a neural network trained on an external dataset but with slice level labels (the dataset in question is a small subset of the training set for the ImageCLEF 2021 - Tuberculosis detection competition \cite{senticlabImageClef2021}, which we manually labeled)
    \item manual correction of the predictions and further labeling 
\end{itemize}

As the COV19-CT-DB dataset it too big to manually address slice labeling, we applied this process to only a small subset of the data. The final dataset we construct will have 3 labels: COVID, non-COVID pneumonia, healthy.

To use the aforementioned neural network trained at ImageClef, we need to apply on the COV19-CT-DB data the same pre-processings used for its training. This turned out to be a complex and imprecise task as the original ImageClef training set used CT volumes in DICOM format and some of the pre-processing steps were applied on the Hounsfield values of the input, while the COV19-CT-DB is in jpg format.

Next, we shortly describe the process involved in building the model for tuberculosis classification in the ImageClef competition, in order to understand its use for building the COVID refined dataset here. Given a selected slice in a CT, we grouped it together with the previous and the next slice in the volume; these mini volumes of 3 consecutive slices, we thought, could better highlight the lesions present in the tuberculosis dataset, emphasising the difference between an infiltration and an artery, or a cavern and a lumen as these can be very similar at a certain point in space but continue in a different manner. We changed its window and level values to highlight the lung features. The selected slices were split into half, corresponding to each lung, and we kept only the side that was labeled. We cropped the images, using a simple threshold method to remove the padding and keep only the body. The resulting images were resized to $256\times256$ pixels. These were then normalized with values in the range $[0,255]$ corresponding to 3 black and white images which were concatenated at channel level.  As augmentations we used a random crop of size $224\times224$, a random horizontal flip with a probability of $0.5$ and normalized the image. We trained an EfficientNet-B3 \cite{tan2019efficientnet}, with batches of 32 for 90 epochs. 

We used this network to predict on each slice of a volume for the COVID patients in the COV19-CT-DB training set 
the presence of an affection, and obtain the probabilities of each affection type that was present in the ImageClef data (a total of five lesion types and one healthy class). 

In order to be able to apply this pre-trained model on the COVID dataset we had to address several problems. The steps we followed to address the pre-processing are:

\begin{itemize}
    \item We aggregated the slices into volumes
    \item Constructed an equivalent cropping functions based on  pixel thresholds
    \item Made a prediction with the pre-trained neural network and filtered the results based on one specific label - "healthy". Using an upper and a lower threshold we filtered out the predictions with a score for the "healthy" label between those values, thus keeping only high confidence predictions. We also filtered based on slice location, preferring slices at the middle at the series so as not to saturate the "healthy" class with slices at the ends of the volume that were by default healthy as they did not contain yet the lungs.
    \item We retrained and made the predictions on the entire original training set. In order to obtain the same number of predictions per patients for the last training we resized the reconstructed volumetric image to a depth of size $96$. As some of the patients' folder have very few slices this means the resize would just multiply the same images along the volume.
    \item For each slice we predicted the probabilities of the 3 classes enumerated above. We grouped the slices predictions into an array of size $(96, 3)$. We did this for each patient, thus creating a data set that was used as training data for a linear classifier that learns to predict the class at CT level.
\end{itemize}

To predict on the validation set we would first have to predict with the first neural network on the reconstructed slices at slice level and then predict on the results with the second classifier at patient level. We trained a logistic regression classifier and a multilayer perceptron with 100 neurons and a single layer. The results on the training and validation sets are shown in tables 2 and 3.

\section{A slice based approach tackled with neural network architecture search}\label{sec:NNSearch}

This approach is intended to act as our own baseline, as it consists of very simple steps that could be taken to build a classifier for CT data:
\begin{itemize}
    \item label each CT slice in the training set with its corresponding CT label;
    \item use an existing implementation that performs neural architecture search for image classification to generate a network architecture that performs well on our problem at hand - classifying the CT slices;
    \item aggregate the results at slice level using simple statistics in order to classify a CT.
    
\end{itemize}

In the last few years there have been great improvements in the field of Neural Architecture Search (NAS).
The state-of-the-art models are able to find neural network architectures in a small number of GPU days, much less than the long search times that 
were required just a few years ago.
Since we are dealing with a classification problem, employing a neural architecture search algorithm that has a very good result on a benchmark classification dataset is an idea worth exploring.

The current state-of-the-art neural architecture search algorithm for the classification of the cifar-10 dataset is sharpDARTS, 
introduced in \cite{hundt2019sharpdarts}.
The sharpDARTS algorithm operates in such a way that it searches for a primitive that will appear multiple times in a larger neural network architecture.

The authors of the sharpDARTS algorithm are modelling the space of the neural network architectures in such a way that is it differentiable.
In this differentiable search space, gradient based search methods are proven to be very effective.
The authors have also made the implementation of sharpDARTS available \footnote{https://github.com/ahundt/sharpDARTS}.
Starting from the authors' implementation, we have made the necessary adjustments to the implementation in order to run the search on the present dataset.
In this approach, the classification takes place at the level of each CT slice.
We have downsampled the resolution of the images from $512$x$512$ to $192$x$192$ in order to speed up the training process by avoiding the loading in memory of the entire dataset multiple times.

We build the slice-based dataset in a very simple manner, being aware that it has some important flaws: every label in a COVID CT is marked as COVID, while every label in a non-COVID CT is marked as non-COVID. No filtering is performed, meaning that we expect to have many similar slices in the dataset with contradictory labels, coming from both COVID patients and healthy patients, since not all the slices in a COVID CT present COVID specific lesions.

\begin{figure}
    \centering
    \includegraphics[width=8cm]{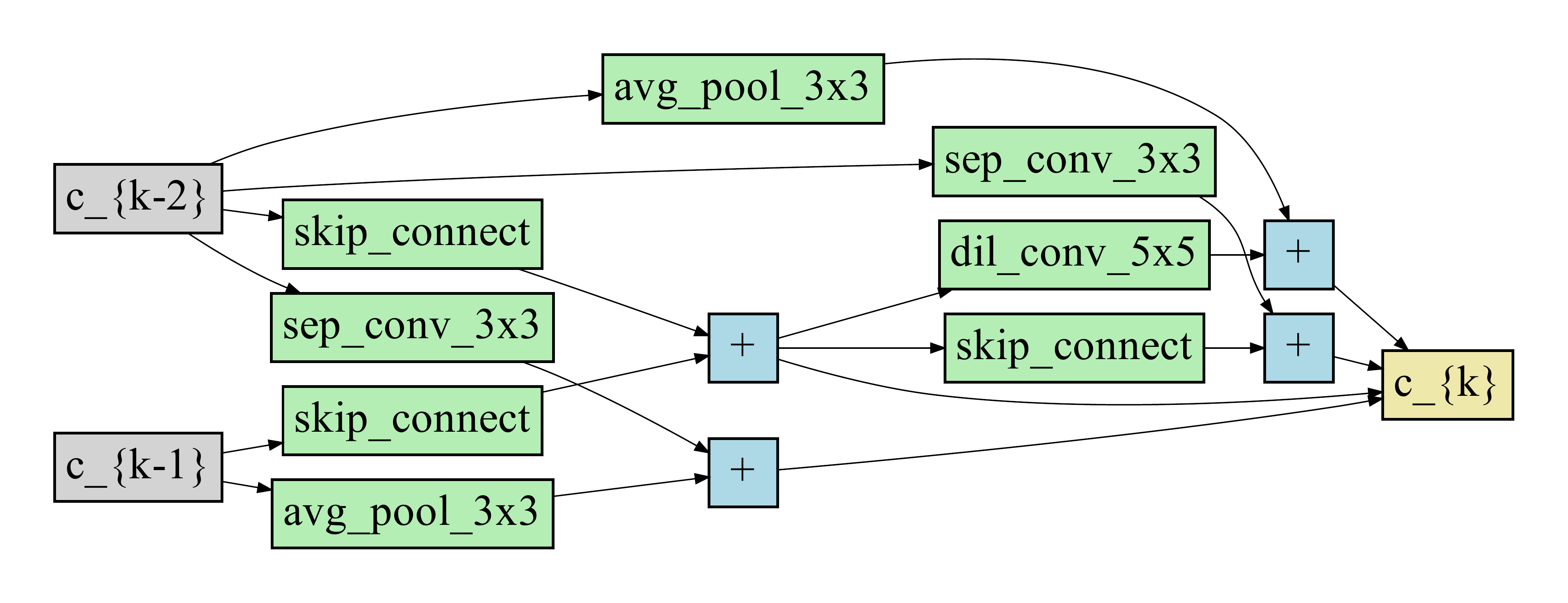} 
    \caption{The primitive found by the sharpDARTS search algorithm.}
    \label{fig:primitive}
\end{figure}

SharpDARTS' authors have found a state-of-the-art model on the cifar-10 dataset in 0.8 days.
We have run the search algorithm for 1 day, finding a primitive presented in Figure \ref{fig:primitive}.

\begin{figure}
    \centering
    \includegraphics[width=8cm]{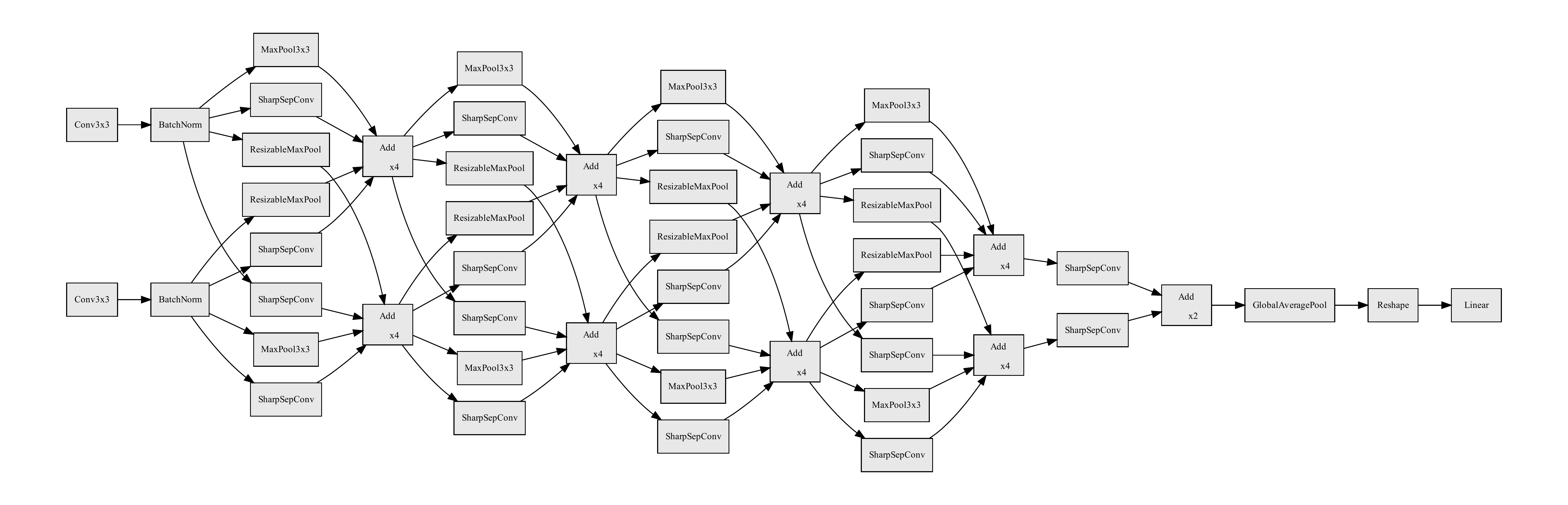} 
    \caption{The final architecture found by the sharpDARTS search algorithm.}
    \label{fig:sharpdarts_arch}
\end{figure}

With this primitive, we have constructed the final architecture as shown in Figure \ref{fig:sharpdarts_arch}.
We continued to train the final neural network architecture for $300$ epochs.

We present the results of this per-slice approach in Table \ref{tab:sharpdarts_slice} - scores are computed here based on slice labels and not CT labels (we measure the performance of the model for labeling slices).

We compute the label for each volume by utilizing the majority vote rule, based on the labels predicted for all the slices in the volume.
The results for CTs classification are presented in Tables  \ref{tab:train} and \ref{tab:vaildation}. Despite it being a relatively small architecture of only $61.3$ MMAC (million multiply-accumulate operations; $122.6$ MFLOPS) and being trained on noisy data, it obtained an unexpectedly good F1 score on the validation dataset of 0.74.

\begin{table}
\begin{tabular}{|l|c|c|c|}
\hline
Data & precision & recall & macro F1\\
\hline
Train & $0.724$ & $0.731$ & $0.727$ \\
Validation & $0.729$ & $0.777$ & $0.752$\\
\hline
\end{tabular}
\caption{Per slice results of the sharpDARTS algorithm}
\label{tab:sharpdarts_slice}
\end{table}

It is worth improving this approach by utilizing the slice labeling procedure described in Section \ref{grouped_slices}.
This would help eliminate the issue of slices that are not informative relative to the covid or non-covid nature of the CT.
We would also like to remind that this approach is at a disadvantage since it only utilizes a single slice at a time, losing context information.

\section{Experimental results}\label{sec:experiments}

Tables \ref{tab:train} and \ref{tab:vaildation} present the results obtained on the training and validation sets, in terms of precision and recall for the COVID class, and the macro F1 score, as required in the competition.

All the approaches taken outperform the baseline score set by the competition organizers \cite{kollias2021mia}.

The volumetric approach achieves the best results, bringing a consistent improvement over the baseline.

Although with lower accuracy, we consider the slice-based approaches to be more relevant in practice, providing more information related to lesion localization in the lungs and implicitly its extension/size. Given our previous experience on similar CT classification tasks \cite{senticlabImageClef2020,senticlabImageClef2021}, we argue that the poorer results are due to the presence of noisy labels at slice level in the training data (because of the automated slice labeling procedures we use) and filtering manually COVID slices is necessary to build a good training set for increasing the performance of this approach.

\begin{table*}[h]
\begin{center}
\begin{tabular}{|l|c|c|c|}
\hline
Method & precision (Covid class) & recall (Covid class) & macro F1 score\\
\hline\hline
volumetric approach  & $0.97$ & $0.96$ & $0.96$\\
slice-wise approach based on minivolumes \& log reg & $0.95$ & $0.89$ & $0.93$ \\
slice-wise approach based on minivolumes \& mlp & $0.98$ & $0.93$ & $0.97$ \\
slice-wise approach based on sharpDARTS & $0.93$ & $0.94$ & $0.93$ \\
\hline
\end{tabular}
\end{center}
\caption{Results on the train set for: the volumetric approach described in Section \ref{sec:volumetric}, the two slice based approaches based on mini-volumes described in Section \ref{grouped_slices} and for the slice based approach using neural architecture search described in Section \ref{sec:NNSearch}}
\label{tab:train}
\end{table*}

\begin{table*}[h]
\begin{center}
\begin{tabular}{|l|c|c|c|}
\hline
Method & precision (Covid class) & recall (Covid class) & macro F1 score\\
\hline\hline
volumetric approach & $0.95$ & $0.88$ & $0.92$ \\
slice-wise approach based on minivolumes \& log reg & $0.83$ & $0.71$ & $0.82$ \\
slice-wise approach based on minivolumes \& mlp & $0.87$ & $0.76$ & $0.84$ \\
slice-wise approach based on sharpDARTS & $0.67$ & $0.82$ & $0.74$\\

\hline
\end{tabular}
\end{center}
\caption{Results on the validation set for: the volumetric approach described in Section \ref{sec:volumetric}, the two slice based approaches based on mini-volumes described in Section \ref{grouped_slices} and for the slice based approach using neural architecture search described in Section \ref{sec:NNSearch}}
\label{tab:vaildation}
\end{table*}

\section{Conclusions}\label{sec:conclusions}
The paper presents the approaches taken by the SenticLab.UAIC team for the problem of COVID19 detection in CTs. Given the nature of the data, consisting of volumetric (3D) images, two distinct ways to perform CT classification are used: one that treats the CT as a whole, and one that performs classification at slice level and then aggregates the results at CT level. The experiments show very promising results, reaching a 0.92 macro F1 score.

{\small
\bibliographystyle{ieee_fullname}
\bibliography{egpaper_final}

\begin{thebibliography}{10}\itemsep=-1pt

\bibitem{inflated2017}
Jo{\~{a}}o Carreira and Andrew Zisserman.
\newblock Quo vadis, action recognition? {A} new model and the kinetics
  dataset.
\newblock {\em CoRR}, abs/1705.07750, 2017.

\bibitem{sam}
Pierre Foret, Ariel Kleiner, Hossein Mobahi, and Behnam Neyshabur.
\newblock Sharpness-aware minimization for efficiently improving
  generalization.
\newblock {\em CoRR}, abs/2010.01412, 2020.

\bibitem{resnet2015}
Kaiming He, Xiangyu Zhang, Shaoqing Ren, and Jian Sun.
\newblock Deep residual learning for image recognition.
\newblock {\em CoRR}, abs/1512.03385, 2015.

\bibitem{hundt2019sharpdarts}
Andrew Hundt, Varun Jain, and Gregory~D. Hager.
\newblock sharpdarts: Faster and more accurate differentiable architecture
  search.
\newblock {\em CoRR}, abs/1903.09900, 2019.

\bibitem{kollias2021mia}
Dimitrios Kollias, Anastasios Arsenos, Levon Soukissian, and Stefanos Kollias.
\newblock Mia-cov19d: Covid-19 detection through 3-d chest ct image analysis.
\newblock {\em arXiv preprint arXiv:2106.07524}, 2021.

\bibitem{kollias2020deep}
Dimitrios Kollias, N Bouas, Y Vlaxos, V Brillakis, M Seferis, Ilianna Kollia,
  Levon Sukissian, James Wingate, and S Kollias.
\newblock Deep transparent prediction through latent representation analysis.
\newblock {\em arXiv preprint arXiv:2009.07044}, 2020.

\bibitem{kollias2018deep}
Dimitrios Kollias, Athanasios Tagaris, Andreas Stafylopatis, Stefanos Kollias,
  and Georgios Tagaris.
\newblock Deep neural architectures for prediction in healthcare.
\newblock {\em Complex \& Intelligent Systems}, 4(2):119--131, 2018.

\bibitem{kollias2020transparent}
Dimitris Kollias, Y Vlaxos, M Seferis, Ilianna Kollia, Levon Sukissian, James
  Wingate, and Stefanos~D Kollias.
\newblock Transparent adaptation in deep medical image diagnosis.
\newblock In {\em TAILOR}, pages 251--267, 2020.

\bibitem{krizhevsky2012imagenet}
Alex Krizhevsky, Ilya Sutskever, and Geoffrey~E Hinton.
\newblock Imagenet classification with deep convolutional neural networks.
\newblock In {\em Advances in neural information processing systems}, pages
  1097--1105, 2012.

\bibitem{senticlabImageClef2020}
Radu Miron, Cosmin Moisii, and Mihaela Breaban.
\newblock Revealing lung affections from cts. {A} comparative analysis of
  various deep learning approaches for dealing with volumetric data.
\newblock In Linda Cappellato, Carsten Eickhoff, Nicola Ferro, and
  Aur{\'{e}}lie N{\'{e}}v{\'{e}}ol, editors, {\em Working Notes of {CLEF} 2020
  - Conference and Labs of the Evaluation Forum, Thessaloniki, Greece,
  September 22-25, 2020}, volume 2696 of {\em {CEUR} Workshop Proceedings}.
  CEUR-WS.org, 2020.

\bibitem{senticlabImageClef2021}
Cosmin Moisii, Radu Miron, and Mihaela Breaban.
\newblock Identifying tuberculosis type in cts.
\newblock In Guglielmo Faggioli, Nicola Ferro, Alexis Joly, Maria Maistro, and
  Florina Piroi, editors, {\em Working Notes of {CLEF} 2021 - Conference and
  Labs of the Evaluation Forum, 2021}, {CEUR} Workshop Proceedings.
  CEUR-WS.org, 2021.

\bibitem{labelsmoothing}
Rafael M{\"{u}}ller, Simon Kornblith, and Geoffrey~E. Hinton.
\newblock When does label smoothing help?
\newblock {\em CoRR}, abs/1906.02629, 2019.

\bibitem{tan2019efficientnet}
Mingxing Tan and Quoc Le.
\newblock Efficientnet: Rethinking model scaling for convolutional neural
  networks.
\newblock In {\em International Conference on Machine Learning}, pages
  6105--6114. PMLR, 2019.

\bibitem{nonlocal2017}
Xiaolong Wang, Ross Girshick, Abhinav Gupta, and Kaiming He.
\newblock Non-local neural networks, 2018.

\end{thebibliography}
}

\end{document}